\documentclass[letterpaper]{article} 
\usepackage{aaai25}  
\usepackage{times}  
\usepackage{helvet}  
\usepackage{courier}  
\usepackage[hyphens]{url}  
\usepackage{graphicx} 
\usepackage{natbib}  
\usepackage{caption} 
\usepackage{algorithm}
\usepackage{algorithmic}

\usepackage{newfloat}
\usepackage{listings}
\DeclareCaptionStyle{ruled}{labelfont=normalfont,labelsep=colon,strut=off} 
\floatstyle{ruled}
\newfloat{listing}{tb}{lst}{}
\floatname{listing}{Listing}
\usepackage{multirow}
\usepackage{xcolor}
\usepackage{booktabs}
\usepackage{amsmath}
\usepackage{amsfonts}
\usepackage{tikz}
\usepackage{subcaption}
\usepackage{pifont}

\DeclareMathOperator*{\argmin}{arg\,min}

\newcommand{\lanyu}[1]{\textcolor{black}{#1}}
\newcommand{\lanyus}[1]{\textcolor{black}{#1}}
\newcommand{\crv}[1]{\textcolor{black}{#1}}

\title{SIDE: Socially Informed Drought Estimation Toward Understanding Societal Impact Dynamics of Environmental Crisis}

\author {
    Lanyu Shang\textsuperscript{\rm 1,3},
    Bozhang Chen\textsuperscript{\rm 1},
    Shiwei Liu\textsuperscript{\rm 1},
    Yang Zhang\textsuperscript{\rm 1},
    Ruohan Zong\textsuperscript{\rm 1},
    Anav Vora\textsuperscript{\rm 2}\\
    Ximing Cai\textsuperscript{\rm 2},
    Na Wei\textsuperscript{\rm 2},
    Dong Wang\textsuperscript{\rm 1}
}
\affiliations {
    \textsuperscript{\rm 1}School of Information Sciences, University of Illinois Urbana-Champaign, Champaign, IL, USA\\
    \textsuperscript{\rm 2}Civil and Environmental Engineering, University of Illinois Urbana-Champaign, Urbana, IL, USA\\
    \textsuperscript{\rm 3}Computer Science, Loyola Marymount UNiversity, Los Angeles, CA, USA\\
    \{lshang3, bozhang4, shiweil2, yzhangnd, rzong2, amvora3, xmcai, nawei2, dwang24\}@illinois.edu
}

\begin{document}
\maketitle
\begin{abstract}
Drought has become a critical global threat with significant societal impact. Existing drought monitoring solutions primarily focus on assessing drought severity using quantitative measurements, overlooking the diverse societal impact of drought from human-centric perspectives. Motivated by the collective intelligence on social media and the computational power of AI, this paper studies a novel problem of socially informed AI-driven drought estimation that aims to leverage social and news media information to jointly estimate drought severity and its societal impact. Two technical challenges exist: 1) How to model the implicit temporal dynamics of drought societal impact. 2) How to capture the social-physical interdependence between the physical drought condition and its societal impact. To address these challenges, we develop SIDE, a socially informed AI-driven drought estimation framework that explicitly quantifies the societal impact of drought and effectively models the social-physical interdependency for joint severity-impact estimation. Experiments on real-world datasets from California and Texas demonstrate SIDE's superior performance compared to state-of-the-art baselines in accurately estimating drought severity and its societal impact. SIDE offers valuable insights for developing human-centric drought mitigation strategies to foster sustainable and resilient communities.

\end{abstract}



\section{Introduction}
\label{sec:intro}

With the increasing frequency and severity, drought has become a critical global threat and poses significant challenges to our societies, leading to detrimental impacts on human well-being, environmental sustainability, and socio-economic development~\cite{sugg2020scoping}. 
For example, drought has cost the agricultural sector of California an estimated \$1.7 billion in direct revenue losses and over 19,000 job losses in 2022~\cite{medellin2022economic}. 
Existing solutions have mainly focused on assessing drought severity with quantitative measurements, such as meteorological and hydrological records,  and agricultural yield data~\cite{dantas2020drought, becker2020never}. However, the underlying \textit{societal impact} of drought (e.g., public health, ecosystem, recreation) has not been systematically explored, especially from human-centric perspectives. Motivated by the collective intelligence on social media in capturing human observations and experiences \lanyus{and the computational power of AI}, this paper focuses on a novel problem of socially informed \lanyus{AI-driven} drought estimation. Our goal is to jointly \lanyus{estimate} drought severity and identify salient determinants of the societal impact precipitated by drought. 

A few recent efforts have been made to investigate the societal impact posed by drought~\cite{sugg2020scoping}. These solutions have mainly focused on scientific measurements, such as meteorological and hydrological data, or economic losses in specific sectors like agriculture~\cite{naumann2021increased}. However, these approaches often overlook the diverse nature of drought's impact on human society, which often extends beyond quantifiable metrics and encompasses various aspects of the human community, such as public health, ecosystem, recreation, and tourism~\cite{savelli2022drought}. Moreover, traditional drought monitoring and assessment methods heavily rely on physical sensors and remote sensing technologies, which may not capture the diverse and localized perceptions of communities affected by drought~\cite{savelli2022drought}. The lack of human-centric perspectives in drought severity assessment can lead to an incomplete understanding of the societal impact and hinder the development of effective mitigation strategies that address the diverse needs and concerns of communities affected by drought.   

\begin{figure}[t!]
    \centering
    \includegraphics[width=\linewidth]{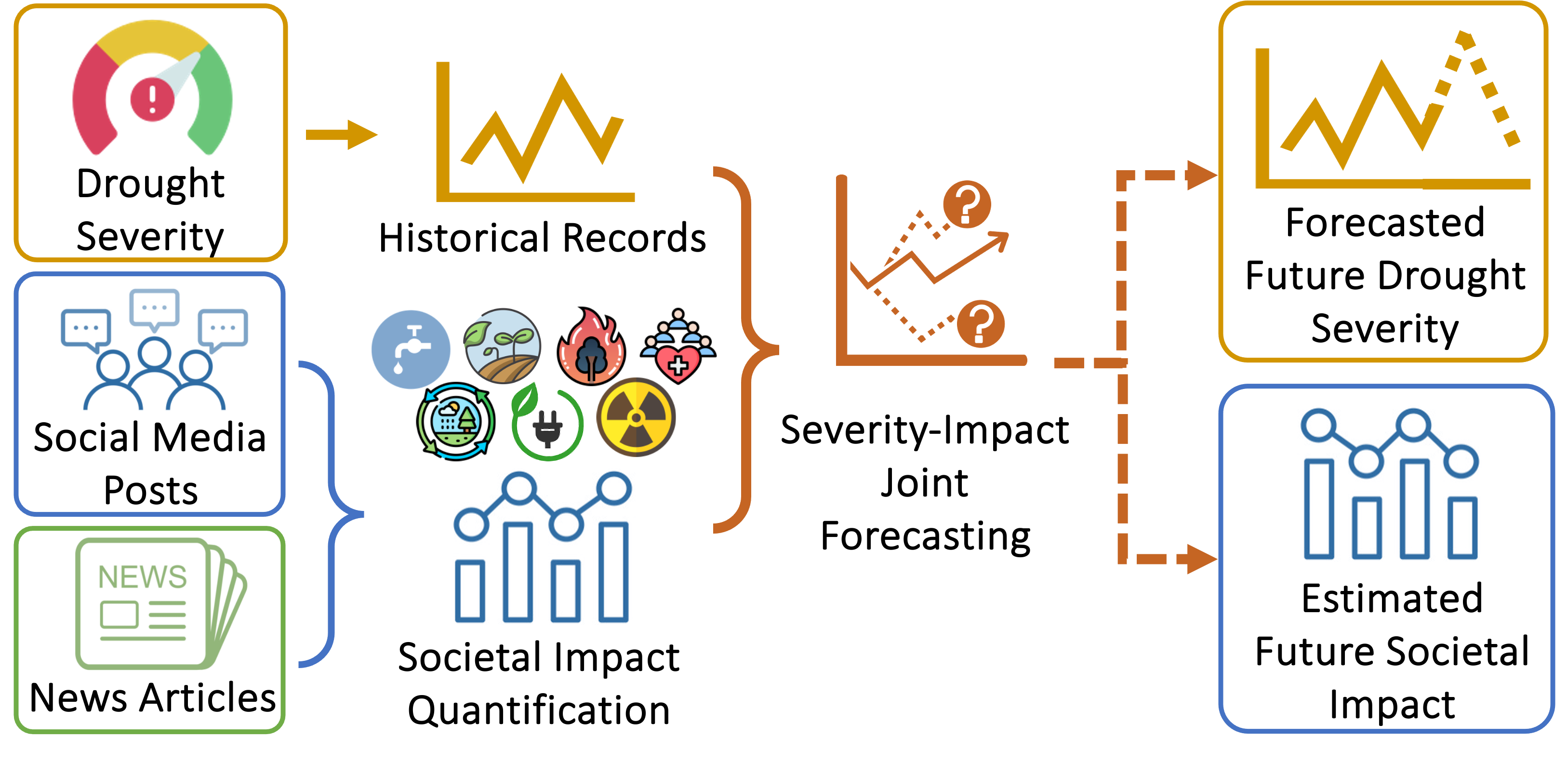}
    \caption{Socially Informed Drought Estimation Overview}
    \label{fig:intro}
\end{figure}

To bridge this gap, in this paper, we study the problem of socially informed \lanyus{AI-driven} drought \lanyus{estimation} by leveraging the collective intelligence on social and news media to \textit{jointly \lanyus{estimate the severity and societal impact of drought}}. Social media platforms and online news outlets have become vital sources of timely, localized, and human-centric information during various natural disasters and crises~\cite{bathaiy2021social}. Figure \ref{fig:intro} shows an overview of the socially informed drought severity \lanyus{estimation} problem. Our proposed framework integrates physical drought severity indicators, social media data, and news media content to provide a comprehensive and human-centric estimation of drought severity and its societal impact. \lanyus{Specifically, while social media inputs reflect the observations and experiences of individual users, news media data offers broader coverage and more neutral perspectives of drought conditions and their consequences analyzed by professional journalists and domain experts.} However, two technical challenges exist in developing such a framework.

\textit{Implicit Temporal Dynamics.}
The first challenge lies in the implicit temporal dynamic of drought's societal impact. The physical conditions of drought dynamically change over time and can be measured with a set of observable indicators (e.g., precipitation, temperature, soil moisture). However, the societal impact of drought often encompasses a spectrum of factors (e.g., the income of farms and small businesses, drought resilience based on the mitigation strategies, physical and mental health of individuals) that are often dynamically changed and cannot be directly assessed. For example, farmers are more concerned about the drought's societal impact on their income during growing seasons as compared to the off-season, when their attention is directed toward strategies for potential drought mitigation. More importantly, the dynamic changes of societal impact often cannot be directly observed or measured in the real world. Therefore, it remains a challenge to accurately capture the dynamic variation of the implicit societal impact to accurately estimate and explain drought's impact on the societal dimension.

\textit{Social-Physical Interdependence.} The second challenge arises from the complex interdependence between the physical condition and societal impact of drought. On one hand, the physical manifestation of drought (e.g., a period of below-normal rainfall or precipitation) can lead to reduced water availability, affecting both surface water and groundwater resources. As a consequence, such water scarcity directly affects agricultural productivity and water quality, resulting in significant societal impact, including reduced income, increased poverty, and food insecurity among these drought-impacted communities. On the other hand, the societal impact subsequently shapes community responses to drought. For example, communities lacking public awareness about drought risks and the importance of water conservation may overconsume water resources in an attempt to mitigate the impact of drought on agriculture, which may further intensify the severity of drought. Therefore, it is crucial to consider the interdependence between the physical condition and societal impact when developing a comprehensive drought \lanyus{estimation} framework.

To address the above challenges, this paper develops SIDE, a socially informed AI-driven drought estimation framework that aims to jointly \lanyus{estimate the drought severity and the societal impact precipitated by drought.} 
To address the first challenge, SIDE explicitly models the dynamics of various determinants of drought's societal impact by exploring drought-related social media discourse and news articles. 
To address the second challenge, SIDE designs a social-physical cross-attention mechanism to effectively capture the interdependence between physical drought conditions and societal responses to accurately \lanyus{estimate} the severity and societal impact of drought. 
Evaluation results demonstrate that SIDE outperforms state-of-the-art baselines in accurately \lanyus{estimating} the drought severity and the societal impact of drought crises. 
\crv{The proposed SIDE framework could be adopted for real-world deployment by integrating with the Dialogue and Information Platform for a Drought Event (DIP-Drought Platform)~\cite{DIP-Drought} developed by our team to provide actionable insights for stakeholders and decision-makers. The platform will leverage SIDE's capabilities to offer real-time drought severity and societal impact estimations, which will allow drought stakeholders (e.g., farmers, business owners, and policymakers) to access timely and comprehensive information about ongoing and potential drought crises.}

\section{Related Work}
\label{sec:related}


\subsection{Societal Impact of Drought}
Traditional approaches for estimating drought's societal impact often rely on professionally collected data, such as hydrological/meteorological measurements~\cite{dantas2020drought}, economic loss~\cite{becker2020never}, surveys/interviews~\cite{edwards2019social}, and literature reviews~\cite{lester2022understanding}. For example, \citet{dantas2020drought} utilized meteorological data (e.g., precipitation and temperature), to assess the severity and spatial extent of drought and link these physical indicators to potential societal impact, such as reduced agricultural productivity and increased food insecurity. \citet{edwards2019social} conducted interviews with farmers to understand the social and emotional impacts (e.g., stress, anxiety, and mental health) of drought during prolonged dry periods. However, these traditional approaches often fall short of capturing the complex and dynamic nature of drought's societal impact due to their narrow focus (e.g., only using physical indicators), limited scalability (e.g., small-scale interviews), and lack of timely information (e.g., previously published studies). To address these limitations, this paper designs a determinant-driven societal impact quantification approach that explicitly explores the timely social and news media data to quantify the societal impact of drought.

\subsection{Time Series Forecasting}
Time series forecasting has been widely studied in various domains, including weather forecasting, stock market prediction, and disease outbreak detection~\cite{lim2021time}. Traditional time series forecasting methods, such as autoregressive integrated moving average (ARIMA)~\cite{shumway2017arima} and exponential smoothing~\cite{hyndman2008forecasting}, have been extensively used for predicting future values based on historical patterns. However, these methods often struggle to capture complex non-linear relationships and long-term dependencies. Recent advancements in deep learning techniques have shown promising results in capturing intricate patterns and dependencies in time series data. For example, \citet{wu2022timesnet} proposed a time-aware graph neural network that incorporates temporal dynamics and graph structure for time series forecasting. \citet{liu2023itransformer} introduced a hierarchical attention mechanism to capture both inter-series and intra-series dependencies for multivariate time series forecasting. However, most existing approaches focus on univariate or multivariate time series data from a single domain which are insufficient to address the socially informed drought estimation problem that integrates multiple heterogeneous data sources (i.e., drought severity, social and news media data). To overcome such a limitation, we propose a social-physical cross-attention mechanism that aims to capture the complex interactions and dependencies between the physical conditions and the societal impact determinants to jointly \lanyus{estimate} the drought severity and its societal impact.

\newtheorem{myDef}{Definition}
\section{Problem Statement}
\label{sec:problem}

\begin{myDef}
\emph{\textbf{Time Step} $t$: the collection period (e.g., weekly) during which the meteorological, social media, and news media data are collected. In particular, we define $T \in \mathbb{Z}^+$ as the total number of time steps in our study and $t$ is the $t^{th}$ time step. 
}
\end{myDef}

\begin{myDef}\label{def:dsci}
\emph{\textbf{Drought Severity} $D_t$: the intensity of drought conditions ($D_t \in \mathbb{R}$) in the studied area at time step $t$. In particular, we adopt the Drought Severity and Coverage Index (DSCI)~\cite{smith2020calibrating} as our primary meteorological metric for evaluating the extent and intensity of drought conditions in the studied areas. \crv{ DSCI is a comprehensive numerical measurement that measures both drought intensity and its impacted area, and has been widely adopted by leading drought monitor agencies (e.g., drought.gov and U.S. Drought Monitor).}
}
\end{myDef}

\begin{myDef} \label{def:social}
\emph{\textbf{Social Input} $\mathcal{S}_t$: a set of drought-related social media posts published during the time step $t$. 
}
\end{myDef}

\begin{myDef}\label{def:news}
\emph{\textbf{News Input} $\mathcal{N}_t$: a set of drought-related new articles collected from leading news publishers during time step $t$.
}
\end{myDef}

\begin{myDef}\label{def:impact}
\emph{\textbf{Societal Impact} $M_t$: the societal impact $M_t$ of drought at time step $t$ is the distribution of a set of $\delta$ determinants specified by the National Integrated Drought Information System~\cite{droughtgov},
including \textit{Agriculture}, \textit{Ecosystems}, \textit{Energy}, \textit{Hazard Planning \& Preparedness}, \textit{Manufacturing}, \textit{Navigation and Transportation}, \textit{Public Health}, \textit{Recreation and Tourism}, \textit{Water Utilities}, \textit{Wildfire Management}, and \textit{Other}. Formally, let $M_t = [m_{t,1}, m_{t,2}, ..., m_{t,\delta}]$
where $m_{t,i} ~\forall~ i \in \{1, 2, \cdots, \delta\} $ represents the impact score for the $i$-th determinant at time step $t$. Each $m_{t,i}$ can be a real number between 0 and 1, where 0 indicates no impact and 1 indicates the most severe impact. We elaborate on the details of quantifying the societal impact from the social and news input in the Solution section.
}
\end{myDef}


\begin{myDef}\label{def:lookback}
\emph{\textbf{Lookback Window} $T_L$: the historical time period consisting of $T_L$, $1 \leq T_L \leq T-1$, time steps within which the input data are used to predict future drought severity and societal impact. $T_L$ is the lookback window size.   
}
\end{myDef}

\begin{myDef}\label{def:prediction}
\emph{\textbf{Prediction Window} $T_P$: the future time period consisting of one or more time steps ($T_P \geq 1$) for which the model predicts drought severity and the societal impact. $T_P$ is the prediction window size.  
}
\end{myDef}

Using the above definition, we formulate the socially informed drought severity \lanyus{estimation} problem as a time series regression problem. In particular, for a time step $t$, given the historical drought severity $D_{t-T_L:t} = [D_{t-T_L}, D_{t-T_L+1}, ..., D_{t}]$, social input $\mathcal{S}_{t-T_L:t} = [\mathcal{S}_{t-T_L}, \mathcal{S}_{t-T_L+1}, ..., \mathcal{S}_{t}]$, and news input $\mathcal{N}_{t-T_L:t} = [\mathcal{N}_{t-T_L}, \mathcal{N}_{t-T_L+1},$
$ \cdots, \mathcal{N}_{t}]$ during the past $T_L$ time steps, our goal is to predict the future drought severity $D_{t+1:t+T_P} = [D_{t+1}, ..., D_{t+T_P}]$ and societal impact $M_{t+1:t+T_P} = [M_{t+1}, ..., M_{t+T_P}]$ for the next $T_P$ time steps. Formally, our problem is formulated as

\begin{equation}
\begin{aligned}
\theta^* = \argmin_{\theta, \varphi} \Bigg(&\frac{1}{T_P} \sum_{i=1}^{T_P} f_\theta(\hat{D}_{t+i}, D_{t+i}) + g_\varphi(\hat{M}_{t+i}, M_{t+i})\\
&\bigm| D_{t-T_L:t}, \mathcal{S}_{t-T_L:t}, \mathcal{N}_{t-T_L:t} \Bigg)
\end{aligned}
\end{equation}
where $f_\theta$ and $g_\varphi$ are the loss function (e.g., mean squared error, mean absolute error) that measures the difference between the predicted drought severity $\hat{D}_{t+i}$ and the actual drought severity $D_{t+i}$, and the predicted societal impact $\hat{M}_{t+i}$ and the actual societal impact $M_{t+i}$, respectively. $\theta$ and $\varphi$ represent the parameters of the \lanyus{estimation} model that need to be optimized. 
\section{Solution}
\label{sec:solution}
\begin{figure}[htb!]
    \centering
    \includegraphics[width=\linewidth]{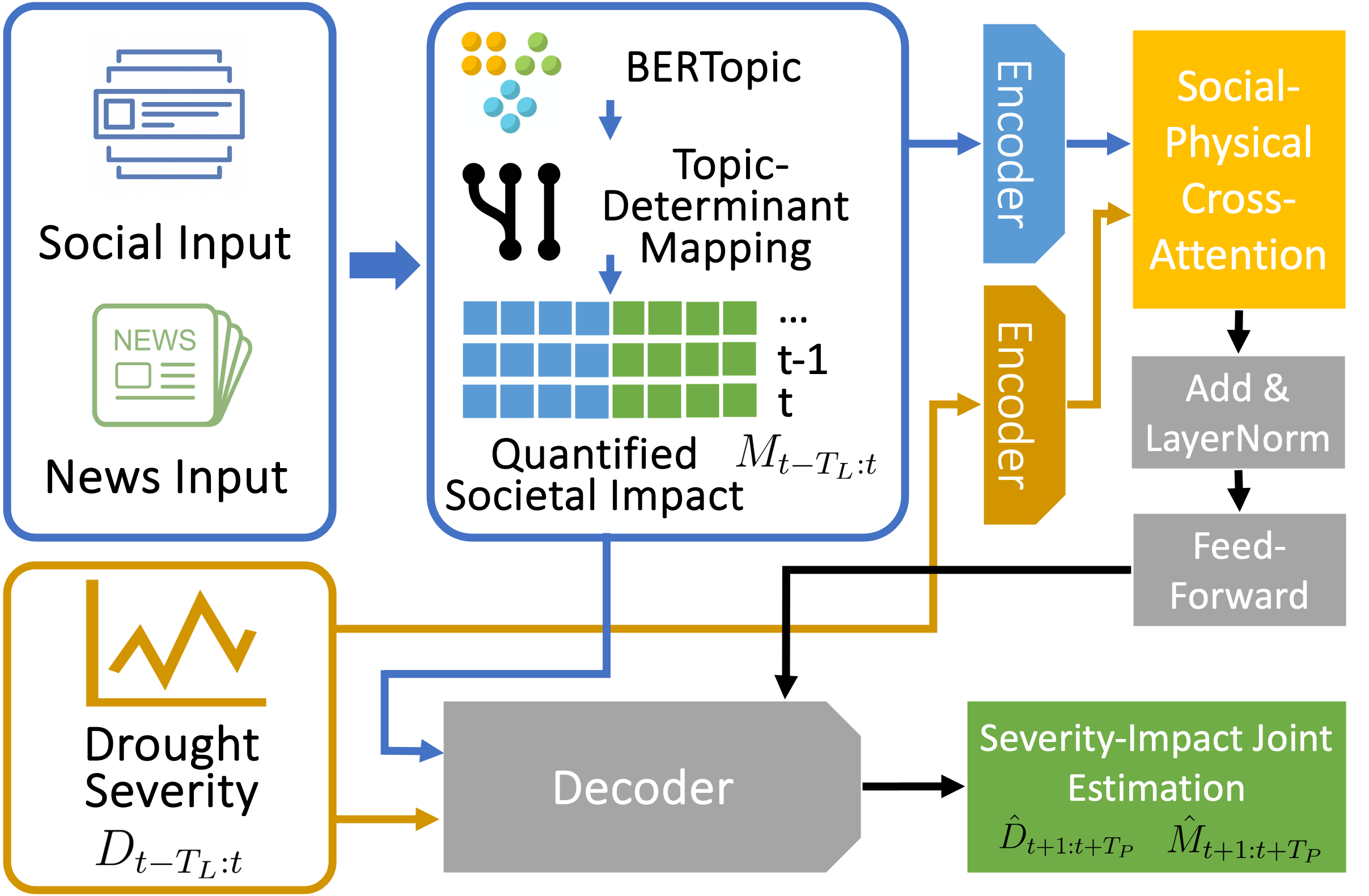}
    \caption{Overview of SIDE}
    \label{fig:solution}
\end{figure}

In this section, we present SIDE, a socially informed drought estimation framework that jointly \lanyus{estimates} drought severity and the underlying societal impact. Figure \ref{fig:solution} shows an overview of the SIDE framework. In particular, SIDE contains three key modules: 1) a \textit{Determinant-Driven Societal Impact Quantification (DSIQ)} module that explicitly quantifies the societal impact of drought by accurately transforming the social and news media data into a comprehensive set of determinants; 2) a \textit{Social-Physical Interdependency Extraction (SPIE)} module that effectively captures the complex interdependency between the distribution of the extracted social determinants and physical drought severity indicator; 3) a \textit{Severity-Impact Joint \lanyus{Estimation} (SIJE)} module that leverages the cross-attended features from SPIE to jointly predict future drought severity and its societal impact.

\subsection{Determinant-Driven Societal Impact Quantification} \label{sol:dsiq}
The determinant-driven societal impact quantification (DSIQ) module is designed to effectively quantify the societal impact of drought by explicitly analyzing social input (Definition \ref{def:social}) and news input (Definition \ref{def:news}). We observe that the social input from social media discourse can provide valuable insights into public sentiment, concerns, and experiences related to drought. Similarly, news input, such as reports from local and national news outlets, can offer a comprehensive overview of the broader societal consequences of drought. However, the social input and news input are primarily unstructured text data, presenting challenges in extracting and quantifying the relevant information for assessing the societal impact of drought. For example, social media posts contributed by individual users often contain a non-trivial amount of noise, such as irrelevant information, opinions, and emotions that may not directly contribute to understanding the societal impact of drought. 

To address such a challenge, we design a hierarchical determinant-driven information distillation strategy to effectively extract drought-related posts/articles from the social and news inputs and quantify their distribution across different societal impact determinants. In particular, the hierarchical determinant-driven information distillation strategy consists of two main steps. First, we extract the fine-grained topics from each social and news input to capture the diverse aspects of drought's societal impact on human communities. We employ the advanced unsupervised neural topic model, BERTopic~\cite{grootendorst2022bertopic}, to discover the latent topics in the noisy text data from social and news inputs. Specifically, considering the varied linguistic characteristics of social media posts and news articles (e.g., language formality, writing style), we train two separate topic models using the social input and news input, denoted as $\Phi_{s}$ and $\Phi_{n}$, respectively. For each topic model, we cluster the corresponding posts/articles into $K$ topic clusters, denoted as $\mathcal{T}_{s}$ and $\mathcal{T}_{n}$.  

Second, we map the extracted topics to a pre-defined set of high-level societal impact determinants (Definition \ref{def:impact}) and learn the distribution of the posts/articles across the determinants to explicitly quantify the societal impact of drought. A straightforward solution is to manually annotate the extracted topics with their corresponding determinants. However, such an approach is not only unscalable but also prone to the dynamic nature of the topics in the social media discourse and news articles as drought condition evolves. To address this challenge, we propose a large language model (LLM) assisted topic-determinant mapping approach that aims to incorporate the commonsense knowledge and natural language understanding capability of LLM to adaptively map the dynamic topics to the relevant societal impact determinants. In particular, we provide the extracted topics and the determinants as prompts and obtain the likelihood score for each topic-determinant pair. Finally, we assign each topic to the determinant with the highest likelihood score and aggregate the posts/articles associated with the topics to compute the distribution of the social media posts and news articles across the determinants. Formally, for a given time step $t$, we define the quantified societal impact as $M_t = [M_{s,t}||M_{n,t}]$
where $M_{s,t}, M_{n,t} \in \mathbb{R}^\delta$ are the normalized distribution of posts/articles across the determinants from the social input and news input, respectively. $\delta$ is the number of determinants. \lanyus{$||$ is the concatenation operation.} The quantified societal impact within the lookback window (Definition \ref{def:lookback}) will be utilized to \lanyus{estimate} the drought severity and predict future societal impact.

\subsection{Social-Physical Interdependency Extraction}
Our next objective is to effectively capture the interdependency between the quantified societal impact and drought severity (Definition \ref{def:dsci}). In particular, we observe that the quantified societal impact and drought severity exhibit complex interactions and dependencies. On one hand, the severity of drought directly influences the societal impact, as more severe drought conditions tend to lead to more significant consequences on various aspects of human society, such as agriculture, economy, and public health. On the other hand, the societal impact of drought can also provide valuable insights and early signals for estimating and predicting drought severity. For example, an increase in social media posts and news articles related to agricultural losses and water scarcity may indicate a worsening drought condition.

To capture such complex interdependency between the quantified societal impact and drought severity, we design a Social-Physical Interdependency Extraction (SPIE) module that designs a social-physical cross-attention mechanism to explicitly learn the hidden relation between the quantified societal impact and drought severity. In particular, we first encode the time series of quantified societal impact $M_{t-T_L:t}$ and drought severity $D_{t-T_L:t}$ within a lookback window of $T_L$ time steps using two separate encoders, $f_{enc}^M$ and $f_{enc}^D$:
\begin{equation}
    H_M = f_{enc}^M(M_{t-T_L:t}), H_D = f_{enc}^D(D_{t-T_L:t})
\end{equation}
where $H_M, H_D \in \mathbb{R}^{T_L \times d}$ are the encoded representations of $M_{t-T_L:t}$ and $D_{t-T_L:t}$, and $d$ is the hidden dimension. \lanyus{$f_{enc}^M$ and $f_{enc}^D$ are the transformer-based sequence encoders for the time series of the societal impact and drought severity, respectively.}

Next, we employ a social-physical cross-attention mechanism to capture the hidden relations between $H_M$ and $H_D$. Specifically, we compute the query, key, and value matrices from $H_M$ and $H_D$ as:
\begin{equation}
    Q_M = H_M W_Q^M;~ K_D = H_D W_K^D;~ V_D = H_D W_V^D
\end{equation}
\begin{equation}
    Q_D = H_D W_Q^D;~ K_M = H_M W_K^M;~ V_M = H_M W_V^M
\end{equation}
where $W_Q^M, W_K^D, W_V^D, W_Q^D, W_K^M, W_V^M \in \mathbb{R}^{d \times d}$ are learnable weights. 
Then, we compute the cross-attention weights $A_{MD}$ and $A_{DM}$ as:
\begin{equation}
    A_{MD} = \text{softmax}(\frac{Q_M K_D^T}{\sqrt{d}});~ 
    A_{DM} = \text{softmax}(\frac{Q_D K_M^T}{\sqrt{d}})
\end{equation}

The cross-attention weights $A_{MD}$ and $A_{DM}$ capture the hidden relations between the encoded representations of quantified societal impact and drought severity. Specifically, each element in $A_{MD}$ represents the attention weight from a time step in $H_M$ to a time step in $H_D$, and vice versa for $A_{DM}$.
Finally, we compute the cross-attended representation of the quantified societal impact $H_{MD}$ and the drought severity $H_{DM}$ as:
\begin{equation}
    H_{MD} = A_{MD} V_D;~ H_{DM} = A_{DM} V_M
\end{equation}

The cross-attended representation $H_{MD}$ and $H_{DM}$ incorporate the hidden relations between the quantified societal impact and drought severity, which are then fed into the subsequent Severity-Impact Joint \lanyus{Estimation} module to enhance the joint prediction of drought severity and societal impact.

\subsection{Severity-Impact Joint \lanyus{Estimation}}
The severity-impact joint \lanyus{estimation} module is designed to jointly predict future drought severity and societal impact by leveraging the cross-attended representation learned from the SPIE module. Traditional time series forecasting models often focus on predicting single or multiple independent target variables based on their historical observations. However, in our problem of socially informed drought estimation, it is crucial to simultaneously estimate both the drought severity and the associated societal impact, as they are inherently interconnected and influence each other dynamically.

To address this challenge, we propose an encoder-decoder architecture that incorporates the cross-attended features from the SPIE module to capture the complex interdependency between drought severity and societal impact. Specifically, let $H_{MD} \in \mathbb{R}^{T_L \times d}$ and $H_{DM} \in \mathbb{R}^{T_L \times d}$ denote the cross-attended representation learned from the SPIE module, where $T_L$ is the lookback window size and $d$ is the hidden dimension. We first concatenate $H_{MD}$ and $H_{DM}$ to obtain a comprehensive representation $H = [H_{MD} || H_{DM}]$ that captures the bidirectional interactions between drought severity and societal impact. 

We then utilize a decoder network $f_{dec}$ to jointly predict the future drought severity and societal impact based on the cross-attended representation $H$. \lanyus{$f_{dec}$ takes $H$ as input and generates the predictions for the next $T_P$ time steps:}
\begin{equation}
    f_{dec}(H) = [\hat{D}_{t+1:t+T_P}, \hat{M}_{t+1:t+T_P}]
\end{equation}
where $\hat{D}_{t+i} \in \mathbb{R}$ and $\hat{M}_{t+i} \in \mathbb{R}^{\delta}$ are the predicted drought severity and quantified societal impact at time step $t+i$, respectively, and $T_P$ is the prediction window size. 

We optimize the SIDE framework by jointly considering the \lanyus{estimation} error for both drought severity and societal impact. Specifically, we define the loss function as:
\begin{equation}
    \mathcal{L} = \sum_{i=1}^{T_P} (\lambda_D \mathcal{L}_D(\hat{D}_{t+i}, D_{t+i}) + \lambda_M \mathcal{L}_M(\hat{M}_{t+i}, M_{t+i}))
\end{equation}
where $\mathcal{L}_D$ and $\mathcal{L}_M$ are the loss functions for drought severity and societal impact, respectively (e.g., mean squared error), and $\lambda_D$ and $\lambda_M$ are hyperparameters that control the relative importance of each task.

By jointly predicting the future drought severity and societal impact, the SIJE module provides more comprehensive \lanyus{estimation} for drought severity and societal impact, aiming to support well-informed decision-making and resource allocation for drought mitigation and management.

\section{Evaluation} \label{sec:eval}
In this section, we evaluate the performance of SIDE on estimating drought severity and its societal impact in different geographical areas.
Evaluation results demonstrate that SIDE achieves substantial performance gains in accurately forecasting the drought severity.

\subsection{Data}
We adopt the publicly available SocialDrought dataset~\cite{shang_2024_10516342} for evaluating the performance of SIDE. In particular, the SocialDrought dataset is a comprehensive drought dataset that contains the meteorological records of drought severity in the U.S. along with the relevant social media discourse \crv{from Twitter/X} and credible news articles \crv{from mainstream news media}. In our experiments, we focus on two states in the U.S., namely California (CA) and Texas (TX), as the primary geographical areas in our study due to their significant agricultural outputs, large and diverse populations, and vulnerability to drought conditions~\cite{becker2020never, farahmand2021drought}. 

We retrieve the weekly DSCI (Definition \ref{def:dsci}), social media posts, and news articles from the SocialDrought dataset for each studied state (i.e., CA and TX) to be used as the \lanyu{ground-truth} drought severity, social input, and news input, respectively. \lanyu{The social and news inputs are further utilized to compute the ground-truth societal impact using the quantification method introduced in the DSIQ module.} In addition, we notice that only a limited amount of the social media posts and news articles in the SocialDrought dataset contain explicit geolocation information that could be used to retrieve relevant posts and articles in each state. To address this issue, we further filter the posts and articles with a list of state-specific location entities to retrieve posts and articles that are relevant to each state. This list of location entities includes major cities, counties, and landmarks specific to California and Texas, \lanyu{such as Los Angeles, Fresno, and Dallas.} This approach allows us to obtain a comprehensive and state-specific dataset for our analysis, ensuring that the social media posts and news articles accurately reflect the drought conditions and public discourse within each state.

In our study, we primarily focus on the period from January 2017 to April 2023 where both California and Texas experienced several significant drought events~\cite{kam2019monitoring, zhang2022quantitative}. For example, California faced severe drought conditions, particularly in 2021, which was recorded as one of the driest years in the state's history~\cite{seager2022mechanisms}. Similarly, Texas encountered extreme drought conditions in 2022, with many parts of the state experiencing exceptional drought, the highest level on the U.S. Drought Monitor scale~\cite{weaver2023case}. 

\begin{figure}[t!]
  \centering
  \begin{subfigure}{0.49\linewidth}
    \centering
    \includegraphics[width=\linewidth]{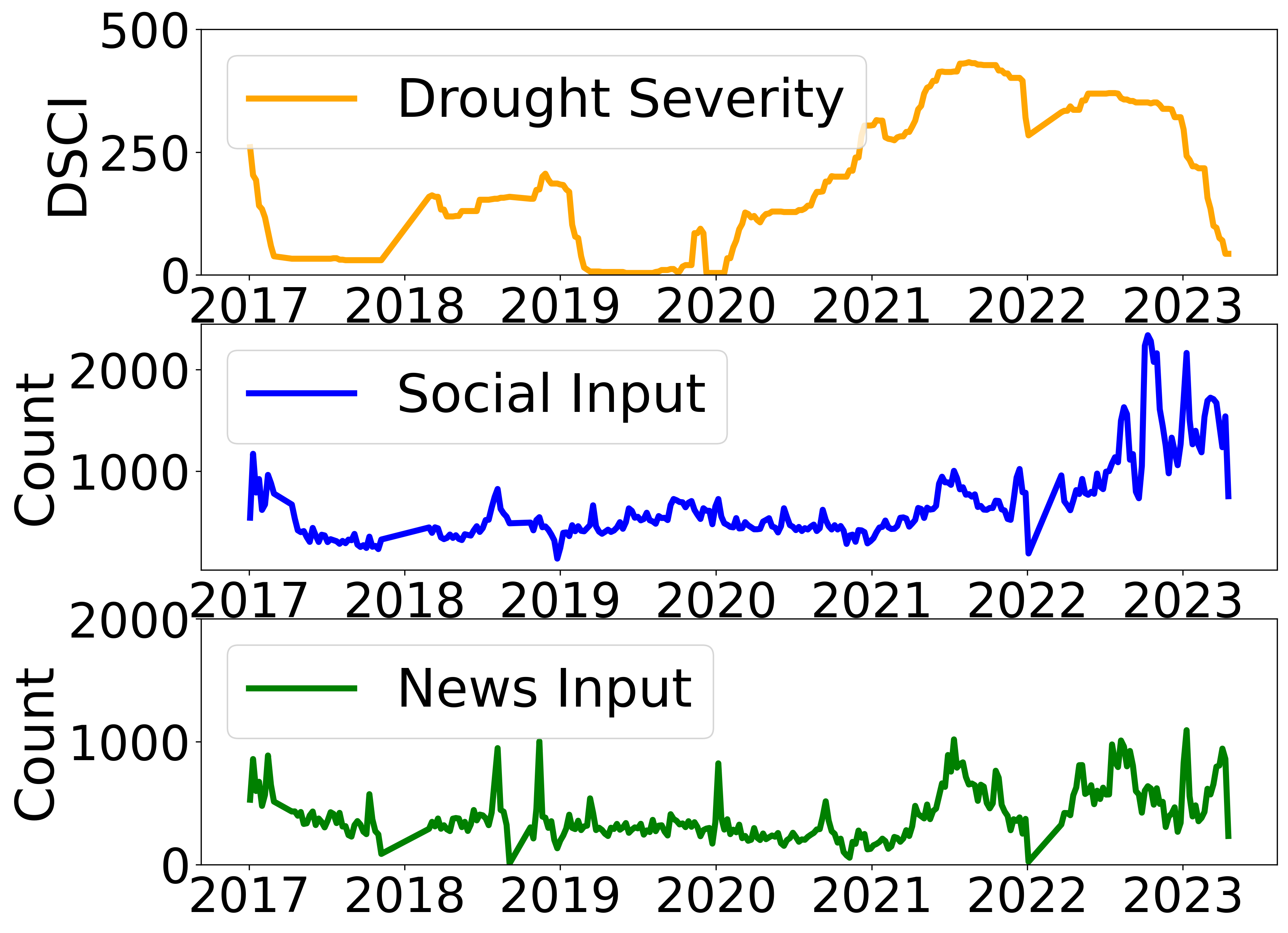}
    \caption{California}
    \label{fig:data_ca}
  \end{subfigure}
  \begin{subfigure}{0.49\linewidth}
    \centering
    \includegraphics[width=\linewidth]{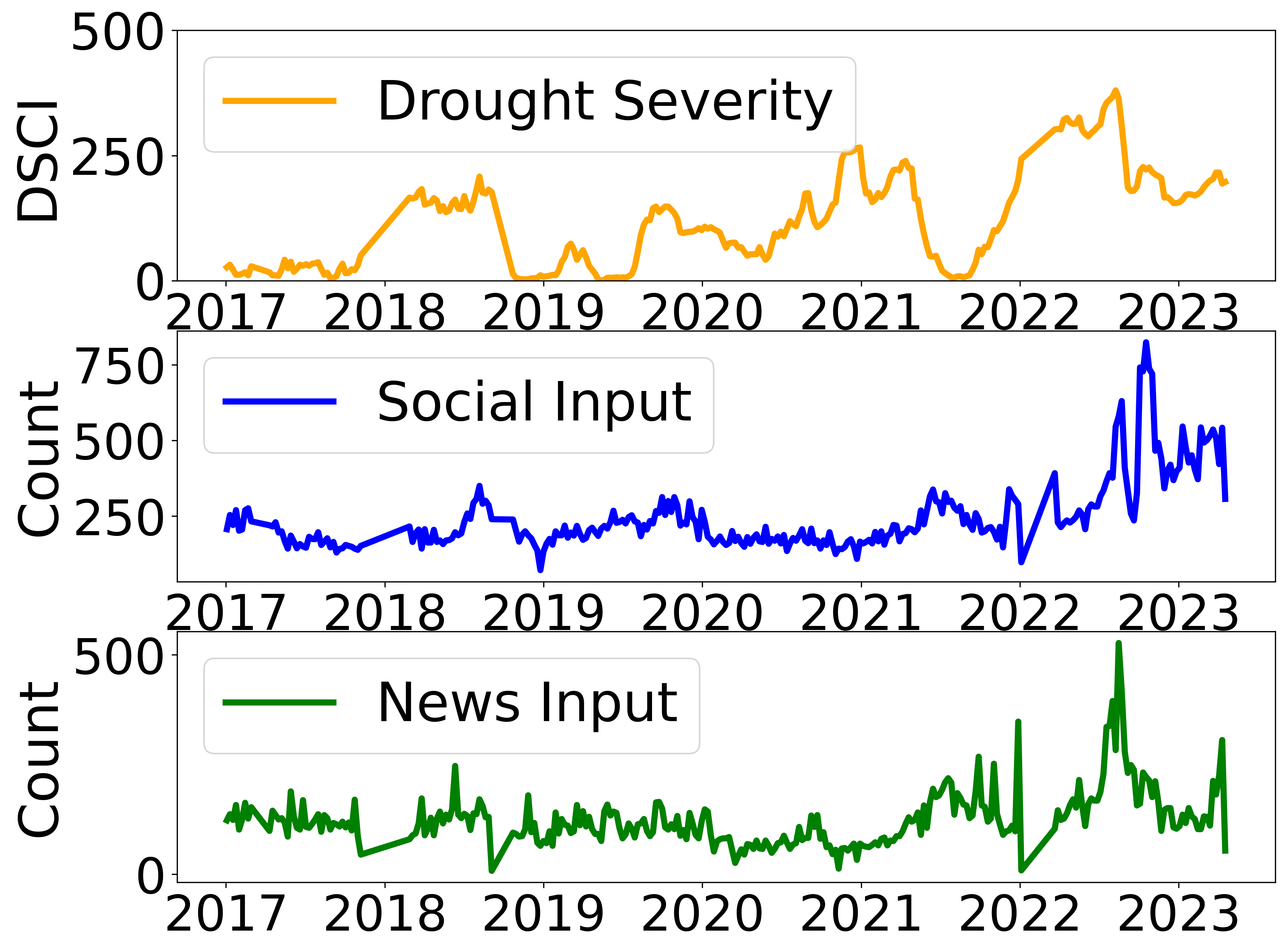}
    \caption{Texas}
    \label{fig:data_tx}
  \end{subfigure}
  \caption{Drought Severity and Social/News Inputs Dynamics}
  \label{fig:data}
\end{figure}

We plot the temporal variation of the DSCI and the amount of weekly social media posts and news articles of California and Texas in Figure \ref{fig:data}. We observe that the temporal patterns of drought severity, social media activity, and news coverage exhibit notable correlations and fluctuations over time. Periods of increased drought severity often coincide with spikes in the volume of drought-related social media posts and news articles in both California (e.g., summer of 2021) and Texas (e.g. summer of 2022). This suggests that social and news media show great potential for providing valuable insights into the societal impact of drought.




\subsection{Baselines and Experimental Settings}

\subsubsection{Baselines}
We compare SIDE with a set of state-of-the-art multivariate time series forecasting solutions. 

\begin{itemize}
    \item \textbf{iTransformer}~\cite{liu2023itransformer}: a transformer-based time series forecasting method that applies the attention and feed-forward network on inverted dimensions.
    
    \item \textbf{Koopa}~\cite{liu2024koopa}: a deep learning approach that utilizes Koopman theory to separate time-variant and time-invariant components in time series forecasting. 
    
    \item \textbf{TimesNet}~\cite{wu2022timesnet}: a neural network architecture that leverages tensor transformation to capture intra- and inter-period variations for time series forecasting. 
    
    \item \textbf{One-Fits-All}~\cite{zhou2024one}: a pre-trained language and vision model with fine-tuning for downstream time series forecasting tasks. 
    
    \item \textbf{RWKV-TS}~\cite{hou2024rwkv}: an RNN-based sequential forecasting approach that is designed to capture the long-term temporal dependencies in time series data. 
    
\end{itemize}

\subsubsection{Experimental Settings}


To ensure a fair comparison, we use the same input in SIDE and all compared baseline models, including the DSCI records, transformed social and news inputs, and their corresponding time steps.  \lanyus{We follow the original implementation of the compared baselines.} We set the lookback window as 52 weeks and the prediction window as 5 weeks. We split each dataset into training, validation, and testing sets using a ratio of 7:1:2. We set $\delta=11$ based on the latest number of societal impact determinants as specified by the National Integrated Drought Information System. We set the number of topics in the DSIQ module as $50$. We train all models using the Adam optimizer with a decay of $0.5$. We run each model with a maximum of 20 epochs and apply early stopping with a patience of 10 epochs to prevent overfitting. We run the experiments on Ubuntu 20.04 with four NVIDIA A16.

\subsection{Drought Severity \lanyus{Estimation} Performance}
To evaluate the drought severity \lanyus{estimation} performance of SIDE, we adopt a series of evaluation metrics that are commonly used for evaluating the performance of time series forecasting, including \textit{Mean Absolute Error (MAE)}, \textit{Mean Squared Error (MSE)}, \textit{Root Mean Squared Error (RMSE)}, and \textit{Median Forecast Accuracy (MFA)}~\cite{chatfield2000time}. A higher value ($\uparrow$) of MFA and lower values ($\downarrow$) of MAE, MSE, and RMSE indicate better \lanyus{estimation} performance. \lanyu{We compare the predicted drought severity with the ground-truth drought severity in the prediction window. }

We summarize the evaluation results for California and Texas in Table \ref{tab:estimation-ca} and Table \ref{tab:estimation-tx}, respectively. We observe that SIDE achieves significant performance gains compared to the baseline methods across all evaluation metrics on both the California and Texas datasets, demonstrating its ability to accurately \lanyus{estimate} drought severity. For example, SIDE outperforms the best-performing baseline (i.e., TimesNet) on the California dataset by 26.25\% and 10.05\% in terms of reducing the MAE and MSE, respectively. We observe similar performance improvements on the Texas dataset. The improvement can be attributed to the social-physical cross-attention mechanism that effectively captures the complex interdependence between dynamic physical conditions and societal impact influencing drought severity.

\begin{table}[htb!]
    \centering
    \footnotesize
    \resizebox{\linewidth}{!}{
    \begin{tabular}{c c c c c }
    \toprule
    \midrule
    Method & MAE ($\downarrow$)  & MSE ($\downarrow$) & RMSE ($\downarrow$) & MFA ($\uparrow$)\\
    \cmidrule{1-5}
    SIDE  &\textbf{30.07} &\textbf{1823.20} &\textbf{42.69} &\textbf{0.94}\\
    \midrule
    iTransformer &49.25 &5913.26 &76.89 &0.93 \\
    \midrule
    Koopa &55.87 &6389.19 &79.93 &0.90 \\
    \midrule
    TimesNet &40.78 &2026.89 &45.02 &0.88 \\
    \midrule
    One-Fits-All &73.67 &26981.50 &164.26 &0.92\\
    \midrule
    RWKV-TS &74.98 &27727.57 &166.51 &0.91\\
    \midrule
    \toprule
    \end{tabular}
    }
    \caption{Drought Severity \lanyus{Estimation} Performance -- CA}
    \label{tab:estimation-ca}
\end{table}

\begin{table}[htb!]
    \centering
    \footnotesize
    \resizebox{\linewidth}{!}{
    \begin{tabular}{c c c c c }
    \toprule
    \midrule
    Method & MAE ($\downarrow$) & MSE ($\downarrow$) & RMSE ($\downarrow$) & MFA ($\uparrow$)\\
    \cmidrule{1-5}
    SIDE  &\textbf{48.08} &\textbf{4369.98} &\textbf{66.10} &\textbf{0.85}  \\
    \midrule
    iTransformer &50.50 &4693.32 &68.50 &0.84 \\
    \midrule
    Koopa &55.87 &6389.19 &79.93 &0.76 \\
    \midrule
    TimesNet &116.06 &17306.74 &131.55 &0.60\\
    \midrule
    One-Fits-All &89.30 &93827.28 &306.31 &0.75 \\
    \midrule
    RWKV-TS &99.63 &101685.62 &318.88 &0.72 \\
    \midrule
    \toprule
    \end{tabular}
    }
    \caption{Drought Severity \lanyus{Estimation} Performance -- TX}
    \label{tab:estimation-tx}
\end{table}

\subsection{Societal Impact Estimation Performance}
\begin{figure}[htb!]
  \centering
  \begin{subfigure}{0.9\linewidth}
    \centering
    \includegraphics[width=\linewidth]{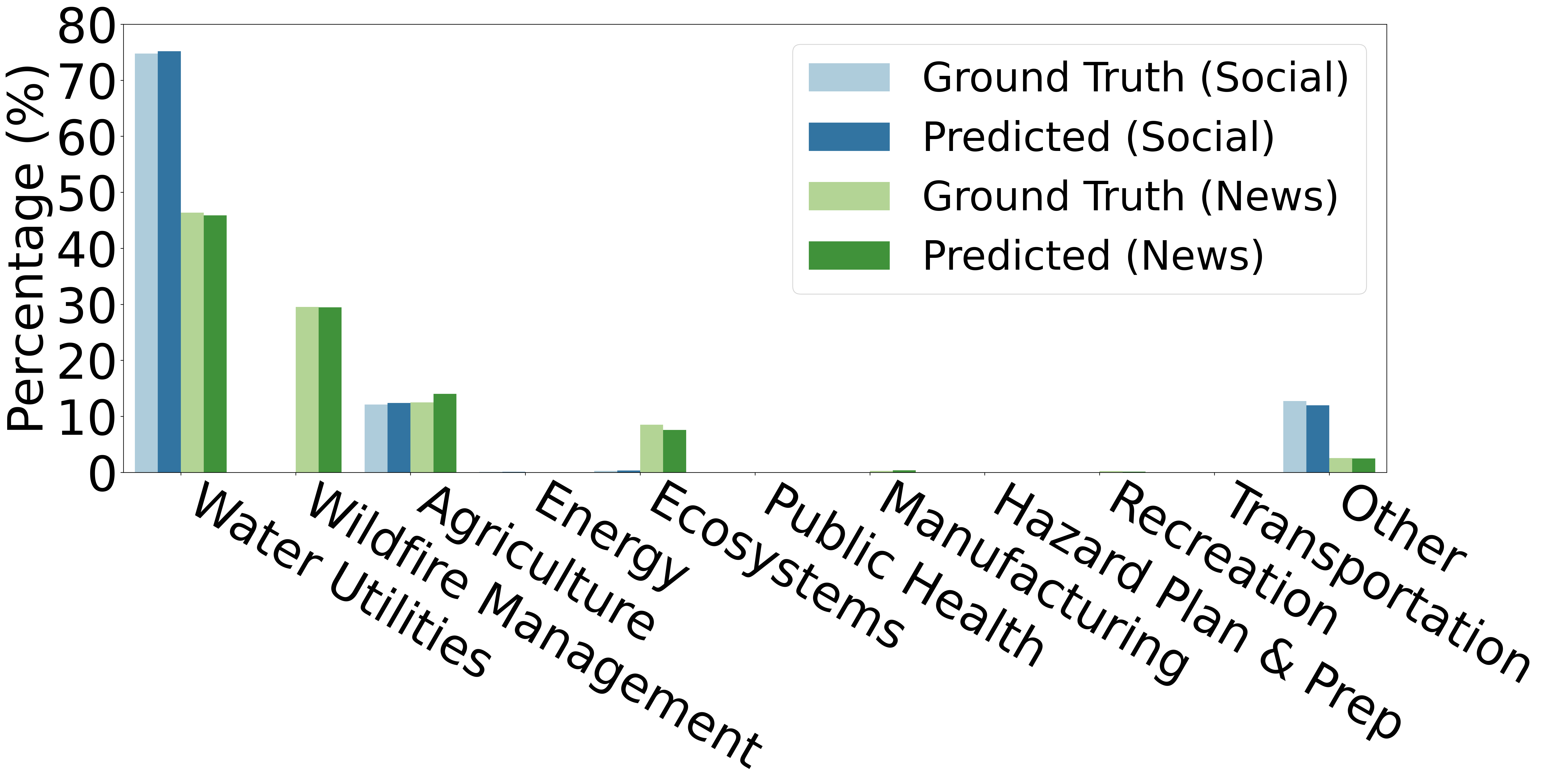}
    \caption{California}
    \label{fig:social_ca}
  \end{subfigure}

  \begin{subfigure}{0.9\linewidth}
    \centering
    \includegraphics[width=\linewidth]{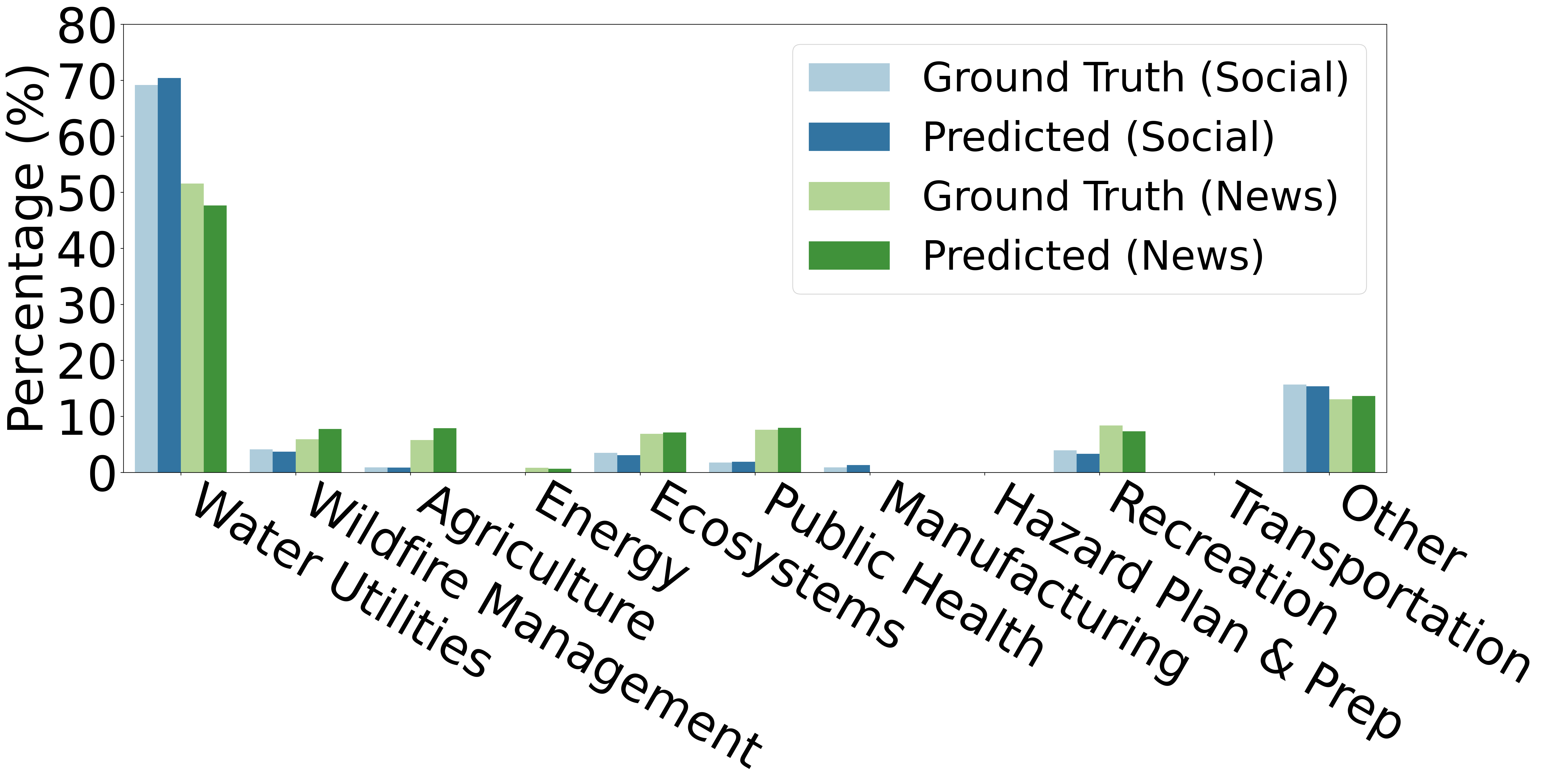}
    \caption{Texas}
    \label{fig:social_tx}
  \end{subfigure}
  
  \caption{Societal Impact Estimation Performance}
  \label{fig:social_est}
\end{figure}

\lanyu{We evaluate the societal impact estimation performance of the SIJE module. In particular, we hold out the societal impact quantified by DSIQ on the prediction window (Definition \ref{def:prediction}), and only use the quantified societal impact from the lookback window (Definition \ref{def:lookback}) as input to SIJE to predict the future societal impact in the prediction window. We compare the societal impact predicted by SIJE with the ground-truth societal impact computed by DSIQ on the prediction window.}
\lanyu{We show the average predicted and ground-truth societal impact of California and Texas in Figure \ref{fig:social_est}.}
We observe that SIJE accurately captures the distribution of the societal impact determinants for both states. 
In addition, we observe that, while ``water utilities'' is the most concerning determinant for both states, California and Texas exhibit distinct patterns in the relative importance of other determinants. For example, ``agriculture'' is shown to be more concerned in California while ``ecosystem'' and ``public health'' are more prominent in Texas. More importantly, we notice that the distribution of the societal impact determinants extracted from the social input and news input also varied (e.g., ``wildfire management'' in California), demonstrating the importance of considering both social and news inputs to comprehensively estimate the societal impact of drought.

\subsection{\crv{Ablation Study}}
We further conduct an ablation study to investigate the contribution of the key components in the SIDE framework. The detailed results and analysis are provided in the Appendix. 

\section{Discussion}
\label{sec:discussion}

\subsection{Deployment}
The proposed SIDE framework can be deployed as a real-time drought monitoring and forecasting system. \lanyus{For example, by integrating SIDE with the U.S. Drought Monitor \cite{droughtmonitor} or DIP-Drought Platform \cite{DIP-Drought} that are developed by leading institutions, we can provide community members with actionable insights for proactive drought management.} 
Moreover, the deployment of SIDE can support various stakeholders, such as government agencies, water resource managers, agricultural organizations, and community leaders. These stakeholders can utilize the timely drought severity \lanyus{estimation} and societal impact assessments provided by SIDE to make informed decisions regarding water allocation, crop management, and the implementation of drought mitigation solutions. For example, water resource managers can use the insights from SIDE to optimize water distribution strategies and prioritize water conservation in areas severely affected by drought. 

\subsection{\lanyus{Broader Social Impact}}
The development of SIDE has broader implications for understanding and addressing the societal impact of environmental crises. By incorporating human-centric perspectives from social and news media, SIDE offers a more comprehensive approach to assessing the diverse impact of drought on human communities. This framework can be adapted to other environmental crises, such as floods, wildfires, and extreme weather events, to gain insights into their societal consequences. Furthermore, the insights derived from SIDE can inform the development of more effective and targeted mitigation strategies and policies that address the specific needs and concerns of affected populations. By fostering a better understanding of the human aspects of environmental crises, SIDE contributes to building resilient and sustainable communities in the face of growing environmental challenges.

\section{Conclusion}
\label{sec:conclusion}
This paper introduces SIDE, a socially informed drought estimation framework that leverages social and news media information to jointly \lanyus{estimate} drought severity and its societal impact. SIDE addresses the challenges of modeling the implicit temporal dynamics of drought societal impact and capturing the social-physical interdependency between drought conditions and societal impact. Extensive experiments on two real-world datasets from California and Texas demonstrate that SIDE substantially outperforms state-of-the-art baselines in accurately \lanyus{estimating} drought severity and its societal impact.

\pagebreak
\section{Acknowledgments}
This research is supported in part by the National Science Foundation under Grant No. CNS-2427070,  IIS-2331069, CHE-2105032, IIS-2130263, CNS-2131622, CNS-2140999. The views and conclusions contained in this document are those of the authors and should not be interpreted as representing the official policies, either expressed or implied, of the U.S. Government. The U.S. Government is authorized to reproduce and distribute reprints for Government purposes notwithstanding any copyright notation here on.

\bibliography{refs}
\section{Appendix}
\subsection{Ablation Study}
\crv{We conduct an ablation study to investigate the individual contributions of the key components in SIDE to the overall drought severity forecasting performance. In particular, we consider the following variants of SIDE: 1) \textbf{SIDE without Social (\textbackslash Social)}: This variant excludes the social input $\mathcal{S}_{t-T_L:t}$ and only uses the historical drought severity $D_{t-T_L:t}$ and news input $\mathcal{N}_{t-T_L:t}$ for forecasting future drought severity. 2) \textbf{SIDE without News (\textbackslash News)}: This variant excludes the news input $\mathcal{N}_{t-T_L:t}$ and only uses the historical drought severity $D_{t-T_L:t}$ and social input $\mathcal{S}_{t-T_L:t}$ for forecasting future drought severity. 3) \textbf{SIDE without Attention (\textbackslash Attn.)}: This variant removes the attention mechanism used for integrating the social and news inputs, and instead directly concatenates the transformed social and news representations with the historical drought severity for forecasting future drought severity. We observe that SIDE achieves its best performance when incorporating all key components.}

\begin{table}[!htb]
    \centering
    \begin{tabular}{c c c c c }
    \toprule
    \midrule
    Method & MAE ($\downarrow$) & MSE ($\downarrow$) & RMSE ($\downarrow$) & MFA ($\uparrow$)\\
    \cmidrule{1-5}
    SIDE  &\textbf{30.07} &\textbf{1823.20} &\textbf{42.69} &\textbf{0.94}\\
    \midrule
    \textbackslash Social &35.28 &2137.63 &46.87 &0.91 \\
    \midrule
    \textbackslash News &33.47 &1926.77 &43.26 &0.92 \\
    \midrule
    \textbackslash Attn. &33.96 &2042.15 &44.05 &0.92 \\
    \midrule
    \toprule
    \end{tabular}
    \caption{Ablation Study -- CA}
    \label{tab:ablation-ca}
\end{table}

\begin{table}[!htb]
    \centering
    \begin{tabular}{c c c c c }
    \toprule
    \midrule
    Method & MAE ($\downarrow$) & MSE ($\downarrow$) & RMSE ($\downarrow$) & MFA ($\uparrow$)\\
    \cmidrule{1-5}
    SIDE  &\textbf{48.08} &\textbf{4369.98} &\textbf{66.10} &\textbf{0.85}  \\
    \midrule
    \textbackslash Social &55.19 &6071.45 &79.03 &0.76 \\
    \midrule
    \textbackslash News &51.97 &5772.86 &72.37 &0.81 \\
    \midrule
    \textbackslash Attn. &50.33 &4517.25 & 68.29 &0.84 \\
    \midrule
    \toprule
    \end{tabular}
    \caption{Ablation Study -- TX}
    \label{tab:ablation-tx}
\end{table}

\end{document}